\documentclass{physeauth}
\usepackage{graphicx}
\usepackage{amsmath}
\usepackage{amssymb}
\usepackage[english]{babel}

\begin{document}

\begin{frontmatter}

\title{
Soft Magnetorotons and Broken-Symmetry States in Bilayer Quantum Hall Ferromagnets
}

\author[SNS,Bell]{Stefano Luin}
\author[SNS]{Vittorio Pellegrini}
\author[Bell,Col]{Aron Pinczuk}
\author[Bell]{Brian S. Dennis}
\author[Bell]{Loren N. Pfeiffer}
\author[Bell]{Ken W. West}

\address[SNS]{NEST-INFM and Scuola Normale Superiore, Piazza dei
Cavalieri 7, I-56126 Pisa (Italy)}
\address[Bell]{Bell Laboratories, Lucent Technologies, Murray Hill,
New Jersey 07974, USA}
\address[Col]{Department of Physics, Department of Applied Physics
and Applied Mathematics, Columbia University, New York, New York
10027, USA}

\begin{abstract}
The recent report on the observation of soft magnetorotons in the
dispersion of charge-density excitations across the tunneling gap
in coupled bilayers at total Landau level filling factor $\nu_T=1$
is reviewed. The inelastic light scattering experiments take
advantage of the breakdown of wave-vector conservation that occurs
under resonant excitation. The results offer evidence that in the
quantum Hall state there is a roton that softens and sharpens
markedly when the phase boundary for transitions to
highly-correlated compressible states is approached. These
findings are interpreted with Hartree-Fock evaluations of the
dynamic structure factor. The model includes the effect of
disorder in the breakdown of wave-vector conservation and
resonance enhancement profiles within a phenomenological approach.
These results link the softening of magnetorotons to enhanced
excitonic Coulomb interactions in the ferromagnetic bilayers.
\end{abstract}

\end{frontmatter}

\section{Introduction}
Two-dimensional electron systems in quantizing magnetic fields
exhibit a variety of collective phases \cite{Girvinbook}. The
additional degree of freedom associated to layer occupation makes
bilayer systems at total Landau level filling factor $\nu_T=1$
particularly interesting \cite{multihall2}. At this filling factor
a phase transition exists between incompressible and compressible
phases (see Fig.1(c)) \cite{multihall}. The incompressible phases
are stable at large tunnelling gaps $\Delta_{\textrm{\tiny SAS}}$
and/or low interlayer distances d. Compressible phases,
characterized by the absence of the quantum Hall signature,
emerge at large d and low $\Delta_{\textrm{\tiny SAS}}$.

The phase transition is finely tuned by the interplay between
$\Delta_{\textrm{\tiny SAS}}$ with intra- and inter-layer Coulomb
interactions. In the incompressible phases the different
dependences of depolarization and excitonic terms on wave-vector
produce, at  $|{\bf q}|= q > 0$, a magnetoroton (MR) minimum in
the dispersion of the charge-density tunnelling excitation (CDE).
In current theories the incompressible--compressible phase
transition is linked to an instability due to softening of a
magnetoroton of CDE modes\cite{Brey90,Brey93,MacD90}. Recent
experimental results on ground state properties close to
the phase boundary were focused on Coulomb drag and interlayer
tunnelling at very low values of $\Delta_{\textrm{\tiny SAS}}$
\cite{Kell02,Kell03,Spiel00,Spiel01}. These studies suggested the
appearance of a Goldstone mode in the incompressible phase and
offered evidence of superfluid behavior.
\par
Experiments that probe dispersive collective excitations and their
softening as a function of $\Delta_{\textrm{\tiny SAS}}$ and $d$
could provide direct evidence of the impact of excitonic terms of
interactions in the quantum phase transitions of the bilayers at
$\nu_T=1$. In previous experiments, resonant inelastic light scattering
methods were employed to access $q\approx 0$ soft tunneling spin excitations
at even values of $\nu_T$ \cite{PELLprl,PELLScience,PELLSSC}.
\par
In the work reported here resonant inelastic light scattering is
employed to probe the low-energy charge excitations of bilayer quantum
Hall (QH) ferromagnets at $\nu _T$ = 1 in states close to a
compressible-incompressible phase boundary \cite{Luin03}. These
experiments seek direct evidence of soft MR modes with $q \!\sim\!
l_B^{-1}$ ($l_B$ is the magnetic length),
and attempt to uncover the impact of excitonic terms
of interactions in the quantum phase transitions of the $\nu_T=1$
bilayer systems.
\par
\begin{figure}[b]
\begin{center}\leavevmode
\includegraphics[width=1.0\linewidth]{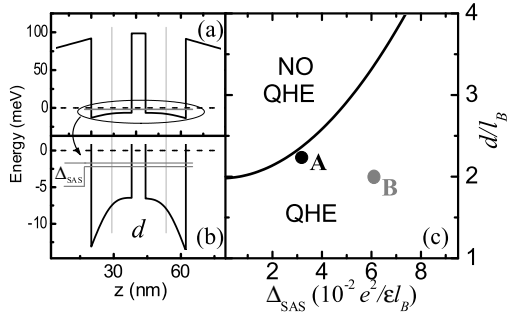}
\caption{(a): Self-consistent calculations of conduction band
diagram of the double quantum well. $d$ is the center-to-center
distance between the two wells. The lowest symmetric and
antisymmetric levels are shown as a dotted horizontal line. (b)
Enlargement of the self-consistent calculation of the conduction
band diagram of the double quantum well close to the tunneling
gap. The dashed horizontal line marks the chemical potential. (c)
Phase diagram of coupled bilayers at $\nu_{T} = 1$. $l_{B}$ is the
magnetic length and $\varepsilon $ the dielectric constant.}
\label{fig1}
\end{center}
\end{figure}
These resonant inelastic light scattering experiments take
advantage of breakdown of wave-vector conservation that occurs in
the QH state due to residual disorder. This aspect of light
scattering methods offers experimental access to critical points
in the mode dispersion such as the one at the MR minimum. To gain
further insights from the data, the results are compared to
time-dependent Hartree-Fock (TDHF) calculations \cite{Wang2002} of
light scattering intensities. The evaluations incorporate effects
of breakdown of wave-vector conservation \cite{marm92},  as well
as the resonant enhancement. This simplified model successfully
reproduces both MR energies and light scattering lineshapes. This
success indicates that the sharpening of the MR spectral lineshape
follows from remarkable changes in the mode dispersion and matrix
elements near the incompressible-compressible phase boundary.
These results uncover significant evidence that softening of
rotons play major roles in the phase transitions of bilayers at
$\nu_T$=1, and suggest a leading role for excitonic Coulomb
interactions in transitions between these highly correlated
phases.

\section{Experimental results}
We present results obtained in two modulation-doped double quantum
wells (DQWs) grown by molecular beam epitaxy. They consist of two
$18\,$nm GaAs wells separated by an undoped Al$_{0.1}$Ga$_{0.9}$As
barrier ($7.5\,$nm for sample A and $6.23\,$nm for sample B).
Figure ~\ref{fig1}(a) shows the self-consistent profile of the
bottom of the conduction band in the DQWs. The dotted line
represent the energy of the lowest symmetric and antisymmetric
states. Their splitting, $\Delta_{\textrm{\tiny SAS}}$ is
below 1 meV as shown in Fig.~\ref{fig1}(b). $d$ is the
center-to-center distance between the two wells.

The samples were designed to have the relatively high
$\Delta_{\textrm{\tiny SAS}}$ of $0.32\,$meV in sample A and
$0.58\,$meV in sample B. Magneto-transport confirms that both
samples are in the quantum Hall side of the $\nu_T=1$
incompressible-compressible phase diagram as shown in
Fig.~\ref{fig1}(c) (filled circles). Detailed analysis of the
magneto-transport features, however, is hindered by the large
parallel conduction found at high magnetic fields. Measured total
sheet densities are $1.2\times10^{11}\,$cm$^{-2}$ in sample A and
$1.1\times10^{11}\,$cm$^{-2}$ in sample B with mobilities larger
than $10^6\,$cm$^2$/V$\,$s. Inelastic light scattering spectra are
obtained in a back-scattering geometry with light propagating
along the magnetic field, almost perpendicular to the 2DEG plane.
Samples are mounted in a $^3$He/$^4$He dilution cryo-magnetic
system with optical windows, at a small tilt angle ($20\,$degrees)
with respect to the incoming laser light. Accessible temperatures
are in the range $50\,$mK$-1.4\,$K. For these measurements the
optical emission of a dye laser is tuned to a frequency $\omega_I$
close to the fundamental interband transitions of the DQW.
Incident power densities are kept below $10^{-4}\,$W/cm$^2$, and
spectra are recorded using a double monochromator, CCD
multichannel detection and spectral resolution of $15\,\mu$eV.

\begin{figure}[b]
\begin{center}\leavevmode
\includegraphics[width=0.75\linewidth]{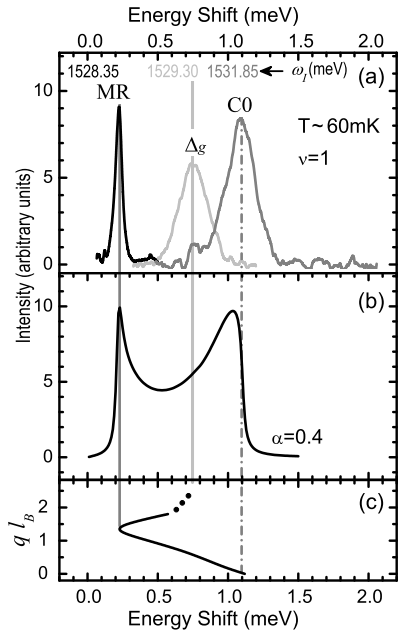}
\caption{(a) Resonant Inelastic light scattering at T=60mK. The
magnetoroton (MR) and $\bf q$$\approx 0$ (C0) and $\bf
q$$\rightarrow \infty$ ($\Delta _g$) charge-density wave peaks are
shown at three different incident photon energies (reported in the
figure). Background due to magneto-luminescence \and laser is
subtracted. (b) Time-dependent Hartree-Fock (TDHF) calculation of
the dynamic strcuture factor as a function of energy for a chosen
value of $\alpha $=0.4 (see text for more details). (c) Collective
mode dispersion within TDHF. Dotted line is the extrapolation at
high values of in-plane wave-vectors.} \label{fig2}
\end{center}
\end{figure}

Figure ~\ref{fig2}(a) shows resonant inelastic light spectra of CDE
modes in sample A with conventional subtraction of the background
due to the laser and to the main magneto-luminescence. The results
are obtained at three different laser wavelengths (indicated in
the upper part of Fig. \ref{fig2}) and display clearly
the three bands of CDE collective modes. In addition to the CDE at
$q\approx 0$ (C0) at $\sim\!\!1.08\,$meV, two
lower-energy excitations are clearly seen. The lowest energy mode
at $\sim\!0.22\,$meV is assigned to the MR critical point in the
dispersion. The MR in Fig.~\ref{fig2}(a) is extremely narrow, with a full width
at half maximum (FWHM) of $\sim\!0.06\,$meV, which is a factor of
three smaller than the FWHM of the C0 band. The peak at
$\sim\!0.75\,$meV, labelled $\Delta_g$, is the large wavevector
CDE excitation. Interpretation of these peaks as the $\Delta_g$ and MR modes
is confirmed by their marked sensitivity on deviations of magnetic
field values from $\nu_T=1$ in a manner that is similar to the QH states.

In the following we further corroborate the assignement of MR peak and
show how the breakdown of wavevector conservation can lead to manifestations of
collective CDE modes at finite ${\bf q}$ in the inelastic light
scattering spectra. To this end we first recall that intensities
of inelastic light scattering is proportional, in first
approximation, to the dynamic structure factor
$S\left(q,\omega;\alpha\right)$. Here $\alpha$ is a
phenomenological broadening parameter in wave-vector space, used to
include the effect of disorder in the breakdown of wavevector
conservation. This model was first introduced by Marmorkos and Das
Sarma \cite{marm92}. Within this model we have:
\begin{equation} \label{eqSS}
S\left(q,\omega;\alpha\right) \sim
\frac{\alpha}{\pi l_B}\int\! dq'
\frac{S\left(q',\omega\right)}{\left(q-q'\right)^2+\left(\alpha\, l_B^{-1}\right)^2}\,,
\end{equation}%
where $S\left(q',\omega\right)$ is the electronic dynamic
structure factor in the translation invariant system.
$S\left(q',\omega\right)$ for charge-density excitations across
the symmetric-antisymmetric gap is given by:

\begin{equation} \label{eqS}
S\left(q,\omega\right) \propto
\frac{\left|M\left(q\right)\right|^2\omega_C\!\left(q\right)\,\omega\Gamma}
{\left[\omega^2-\omega_C^2\!\left(q\right)\right]^2+\omega^2\Gamma^2}\,,
\end{equation}%
where $\Gamma$ is a homogeneous broadening, $\omega _{C}(q)$ the
dispersion of tunneling CDE and $|M(q)|^2$ is a matrix element that acts as an
oscillator strength for inelastic light scattering by collective
excitations. Figure ~\ref{fig2}(b) shows the calculated spectra at
$\alpha$=0.4. This value yields an effective length scale of
$3\cdot l_B$ for the disorder potential involved in the breakdown
of wave-vector conservation at $\nu _{T}$ = 1. The calculated CDE
mode dispersion is shown in Fig.~\ref{fig2}(c).  As expected the
calculated dispersion displays a deep magneto-roton. The predicted
energies of the CDE at $q=0$ (C0) and the MR are consistent with
the experimental results. The evolution of this soft MR with changes in
density, $\Delta_{\textrm{\tiny SAS}}$, and $d$ can unravel the nature
of the phase transition at $\nu _T$ = 1 and will be the subject
of future works. We note that the TDHFA being a mean-field theory fails to reproduce the
collective mode dispersion when $q\gg l^{-1}_B$. That part of the
dispersion (dotted line in Fig.~\ref{fig2}(c)) was not included in
the calculations.

\begin{figure}[bt]
\begin{center}\leavevmode
\includegraphics[width=0.75\linewidth]{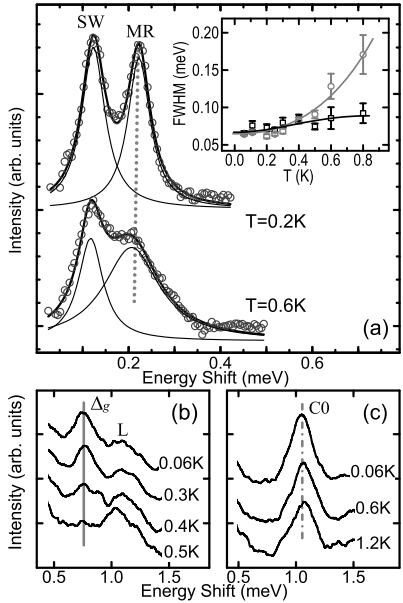}
\caption{(a) Open circles show the temperature
dependence of inelastic light scattering spectra in sample A
after background subtraction. The solid
lines are results of fits with two lorentzians. Inset: temperature
dependence of the full width at half maximum (FWHM) for the spin-wave (SW) peaks
at the Zeeman (black empty squares) and the magnetoroton (MR) peaks (gray empty
circles). The solid lines are guides for the eyes and the error
bars are standard deviations for results on different measurements
and with different background subtractions. (b) Temperature dependence of
the $q\rightarrow\infty$ excitation $\Delta _{g}$; $L$ labels a
magneto-luminescence peak. (c) Temperature dependence of
the $q=0$ tunneling excitation $C0$.}
\label{fig3}
\end{center}
\end{figure}

The temperature dependences of the three excitations shown in
Fig.~\ref{fig2} are reported in Fig.~\ref{fig3}. Panel (a) of
Fig.~\ref{fig3} shows two representative spectra in which both
spin-wave (SW) at the Zeeman gap and MR mode are observed. The
linewidth of the two excitations have very different temperature
dependence as shown in the inset to Fig.~\ref{fig3}(a). FWHM
values are obtained by Lorentzian lineshape fits [solid lines in
Fig.~\ref{fig3}(a)].

We note that for temperatures above 0.8K the MR mode can no longer
be observed while minor changes occur in the SW peak. Similar
temperature dependences characterize the $q\rightarrow\infty$
$\Delta _g$ mode (see panel b of Fig.~\ref{fig3}) and the
$\nu_T=1$ QH effect in magneto-transport (not shown). The
characteristic temperature for this behavior is here almost
one-order of magnitude smaller than $\Delta _g$ and more than a
factor of three below the MR energy in sample A. On the other end,
the $q\sim0$ (C0) mode is temperature-independent up to 1.4K (see
panel c of Fig.~\ref{fig3}).

\begin{figure}[tb]
\begin{center}\leavevmode
\includegraphics[width=0.8\linewidth]{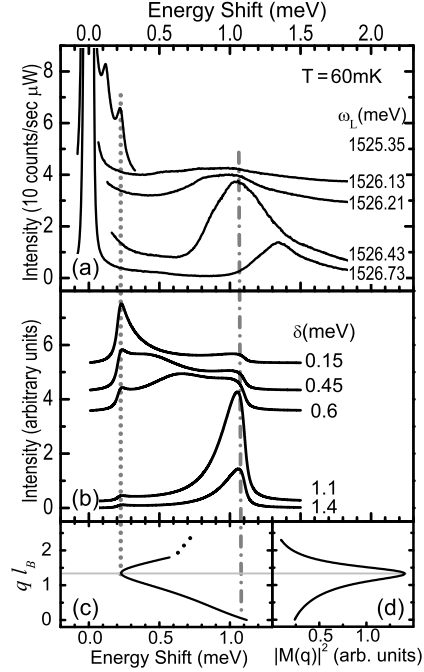}
\caption{(a) Resonant inelastic light scattering spectra of
charge-density excitations (CDEs) in sample A at filling factor
$\nu_T = 1$. The bands were measured with the incident photon
energy $\omega_L$ indicated in meV. (b) Calculated dynamic
structure factor including both breakdown of wave-vector
conservation and resonant enhancement (see text and Eqs. 1,3). (c)
Calculated dispersion of CDE modes. (d) Calculated inelastic light
scattering  ``matrix element''. Vertical lines show the peak
position of CDE modes C0 and MR. The horizontal line in (c) and
(d) marks the MR wave vector.} \label{fig4}
\end{center}
\end{figure}

\begin{figure}[btp]
\begin{center}\leavevmode
\includegraphics[width=0.8\linewidth]{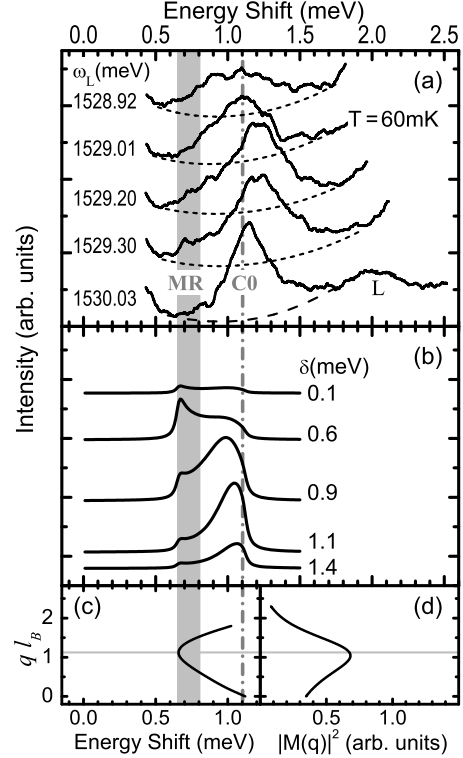}
\caption{Same as in Fig. \ref{fig4} but for sample B.
(a) Resonant inelastic light scattering
spectra of CDEs in sample B at $\nu_T = 1$. The bands were
measured with the incident photon energy $\omega_L$ indicated in
meV. (b) Calculated dynamic structure factor including both breakdown of
wavevector conservation and resonant enhancement (see text and Eq. \ref{eqS}).
(c) Calculated dispersion of CDE modes. (d) Calculated inelastic light
scattering  ``matrix element''. Vertical lines show the
peak position of CDE modes C0 and MR. The horizontal line in (c) and (d) is
at the MR wave vector.}
\label{fig5}
\end{center}
\end{figure}

Figure~\ref{fig4} reports a detailed comparison between
experimental inelastic light scattering spectra of sample A as a
function of laser frequency (a) and calculated spectra that
include the effect of the resonant enhancement profile (b). The
latter can be modelled by replacing the dynamic structure factor
in (\ref{eqS}) with:

\begin{eqnarray}
&&S\left(q,\omega;\delta\right)\propto  \nonumber\\
&&\frac{\left|M\left(q\right)\right|^2\omega_C\!\left(q\right)\,\omega\,\Gamma\,\gamma}
{\left\{\left[\omega^2-\omega_C^2\!\left(q\right)\right]^2+\omega^2\Gamma^2\right\}
\left[\gamma^2+\left(\delta-\omega_C\!\left(q\right)\right)^2\right]}\,,\label{eqS1}
\end{eqnarray}%
where $\delta$ is the position and $\gamma$ is the width of the
resonance. These two parameters can be related to the energy shift
from the laser and the width of the luminescence caused by the
same interband transition responsible for the outgoing resonant
enhancement in the light scattering process. In order to
facilitate the comparison between experiment and theory we plot in
panel (c) the energy dispersion of the CDE mode shown in
Fig.\ref{fig2}. Panel (d) of Fig.~\ref{fig4} displays the matrix
element $|M(q)|^2$ that enters the dynamic structure factor. Large
value of $|M(q)|^2$ at MR wavevector, compared to that of sample B
and shown in Fig. \ref{fig5}, and its sharpness explain the
significant manifestations of MRs in the light scattering spectra
of sample A.

In order to emphasize the connection between the soft and sharp MR
modes observed in sample A and the instability of incompressible
quantum Hall state, we studied a second sample far from the phase
boundary (sample B in Fig.~\ref{fig1}(c)). Figure~\ref{fig5} (a)
shows the light scattering spectra observed in sample B. The C0
band has similar energy and width of that of sample A. It has, in
addition, a marginal field dependence and also occur in spectra
obtained at B=0. The manifestation of the MR structure, on the
other end, is remarkably different from what observed in sample A:
it appears as a weak shoulder with a cutoff at $0.65\,$meV. The
resonant enhancement profile measured in sample B and shown in
Fig. \ref{fig5}(a) reveals a characteristic outgoing resonance
with the higher optical interband transition of the luminescence
peak labelled L.

The TDHF calculations of the dynamic structure factor (Eqs.
\ref{eqSS} and \ref{eqS1}) and of the mode dispersion and
oscillator strength are reported in Fig. \ref{fig5} panels b,c,d,
respectively. The relatively broad and less intense structure of
$|M(q)|^2$ explains the weaker manifestation of MRs in the spectra
of sample B. These results thus offer the framework to understand
energy position, width and relative intensities of the collective
modes observed in the resonant inelastic light scattering
experiments in the two samples.

\section{Discussion and Conclusion}
Our experiments succeeded in detecting the low-lying magnetoroton
minimum in the electron bilayers in AlGaAs/GaAs double quantum
wells at total filling factor $\nu _T$ =1.
These results are intriguing and show that the MR peak softens and sharpens
markedly when the phase boundary for transitions to highly
correlated compressible states is approached.
The observed different temperature-dependence behavior of SW and MR
excitations shown in Fig.\ref{fig3} is also intriguing. In previous
works \cite{Lay94,Abolf00} an anomalous behavior in thermally-activated
magneto-transport was interpreted as evidence for a
finite-temperature transition towards an uncorrelated state. The
evolution of the MR linewidth shown in the inset of
Fig.~\ref{fig3}(a) supports this conclusion. From the smooth
increase of the MR linewidth with temperature we may infer that
thermal fluctuations could destroy the incompressible state and
trigger a transition at finite temperature. It is worth noting
that a similar temperature behavior
characterizes roton excitations in superfluid $^4$He \cite{Glyde90,Svens96}.

In this paper we also provided a detailed description
of resonant enhancement profiles and collective-mode lineshapes
in the resonant inelastic light scattering spectra based on a phenomenological model
that incorporates the disorder-induced breakdown of wave-vector conservation
within a time-dependent Hartree-Fock approach.

The results of the experiments discussed in this paper
confirm that soft rotons play major roles in the
incompressible-compressible phase transition and suggest a leading role
of excitonic Coulomb interactions in these transitions \cite{Brey90,MacD90,Luin03,Wang2002}.
The results support the picture that the ground state of the
bilayers at $\nu_T$=1 evolves towards a broken-symmetry state
caused by the excitonic collapse of the energy of tunneling
excitations at finite wave-vector\cite{Brey90,MacD90}.
The marked narrowing of the MR band
and its interpretation within a time-dependent Hartree-Fock approximation
suggest that the new ground state might be characterized
by a roton wave-vector $q_{R}\sim\!{l_B}^{-1}$.

The results presented here create
venues for further light-scattering experiments at $\nu _{T}$ = 1
that seek definite links between soft modes with the
incompressible-compressible phase boundary.

\section{Acknowledgements}
We are grateful to S. Das Sarma, E. Demler, S.M.
Girvin, A.h. MacDonald, S.H. Simon and D.W. Wang for significant discussions.
This work was supported by the Nanoscale
Science and Engineering Initiative of the National Science
Foundation under NSF Award Number CHE-0117752, by a research grant
of the W. M. Keck Foundation, by CNR (Consiglio Nazionale delle
Ricerche), by INFM/E (Istituto Nazionale per la Fisica della
Materia, section E), and by the Italian Ministry of
University and Research (MIUR)

\bibliography{Luin}




\end{document}